\documentclass[letterpaper,aps,prb,twocolumn,showpacs]{revtex4}

\usepackage{graphicx}
\usepackage{amsmath}
\usepackage{color}
\usepackage{epstopdf}
\begin{document}

\title{Theoretical study of optical conductivity of graphene with magnetic and nonmagnetic adatoms}

\author{Muhammad Aziz Majidi$^{1,2,3}$}
\email{aziz.majidi@sci.ui.ac.id}
\author{Syahril Siregar$^{3}$}
\author{Andrivo Rusydi$^{1,2}$}
\email{phyandri@nus.edu.sg}
\affiliation{$^{1}$ NUSNNI-NanoCore, Department of Physics, Faculty of Science,
National University of Singapore, Singapore 117542, Singapore,}
\affiliation{$^{2}$ Singapore Synchrotron Light Source, National University of Singapore, Singapore 117603,
Singapore}
\affiliation{$^{3}$ Departemen Fisika, FMIPA, Universitas Indonesia,
Depok 16424, Indonesia}
\date{\today}

\begin{abstract}
We present a theoretical study of the optical conductivity of graphene with magnetic and nonmagnetic adatoms. First, by introducing alternating potential in a pure graphene, we demonstrate a gap formation in the density of states and the corresponding optical conductivity. We highlight the distinction between such a gap formation and the so-called Pauli blocking effect. Next, we apply this idea to graphene with adatoms by introducing magnetic interactions between the carrier spins and the spins of the adatoms. Exploring various possible ground-state spin configurations of the adatoms, we find that antiferromagnetic configuration yields the lowest total electronic energy, and is the only configuration that forms a gap.
Furthermore, we analyze four different circumstances leading to similar gaplike structures and propose a means to interpret the magneticity and the possible orderings of the adatoms on graphene solely from the optical conductivity data. We apply this analysis to the recently reported experimental data of oxygenated graphene.
\end{abstract}


\maketitle

\begin{center}
\textbf{I. INTRODUCTION}
\end{center}

Setting adatoms on graphene has been a promising means to tailor electronic properties of graphene, such as transport, optical, and magnetic properties \cite{Schedin_Nat_Mat_2007,Chen_Nat_Phys_2008,Zhou_PRL_2008,Elias_Sci_2009}. In particular, the idea of generating a controllable band gap in graphene while at the same time adding magnetic moments through the adatoms has spawned a broad interest to make graphene a magnetic semiconductor.
Theoretically, it has been predicted that nonmagnetic atoms could become magnetic when put as adatoms on graphene \cite{CastroNeto_SSC_2009}.
An experimental evidence of this is found in fluorinated graphene \cite{Nair_Nat_Phys_2012}.
However, to make the magnetic moments
align to form a ferromagnetic order is still a great challenge to both experimentalists and theorists \cite{Kan_Nano_2008}. Meanwhile, available theories based on Ruderman-Kittel-Kasuya-Yosida (RKKY) mechanism predict that a pair of neighboring magnetic moments of the adatoms on graphene prefer to align antiferromagnetically \cite{Brey_PRL_2007,Saremi_PRB_2007}.
Indication of such antiferromagnetic alignments has been reported recently in fluorinated graphene, for which spin-half paramagnetism is observed with only a very small portion of magnetic moments of the fluorine adatoms contribute to the magnetization \cite{Nair_Nat_Phys_2012}.

In this paper, we aim at showing the intimate connection between magneticity of the adatoms on graphene, their magnetic orders, i.e. antiferromagnetic (AFM), feromagnetic (FM), and paramagnetic (PM), and their corresponding optical responses. In particular, we address how a band gap can form in the density of states (DOS) of this system, and how such a gap reveals in the optical conductivity ($\sigma_1(\omega)$). We analyze several circumstances leading to similar gaplike structures and compare our results with the recently reported experimental data of $\sigma_1(\omega)$ of graphene with oxygen adatoms \cite{Santoso_PRB_2014} at various adatom concentrations.

\begin{center}
\textbf{II. REAL GAP VS PAULI BLOCKING}
\end{center}

\begin{figure}
\includegraphics[width=3.2in]{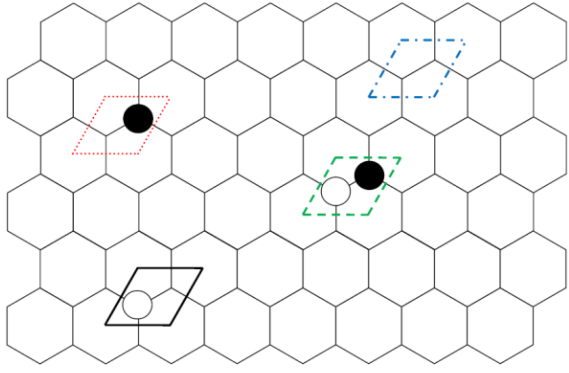}
\caption{(Color online) Illustration of graphene in the presence of adatoms. The honeycomb net represents a graphene lattice. The black filled and opaque circles represent adatoms sitting at sublattice $A$ and $B$, respectively. Parallelograms with different colors and line styles represent supercells with different configurations of adatom occupation: black-solid lines (dotted-red lines) for a supercell containing only an adatom at sublattice $A(B)$, dashed-green lines for a supercell containing adatoms at both sublattices, and dot-dashed-blue lines for a supercell with no adatoms.
}
\label{model_structure}
\end{figure}

Before we go through the more realistic model for graphene with magnetic adatoms, let us consider a toy model of graphene with some artificial potensial $V(-V)$ at each sublattice $A(B)$. Such an alternating potential has been used in the discussion on excitonic mass generation in graphene \cite{Kotov_RMP_2012}. Using the tight-binding approximation defined on a hexagonal lattice as shown as the honeycomb net in Fig. \ref{model_structure}, the Hamiltonian for graphene with alternating potential (neglecting electron-electron interactions) can be written in ${\textbf k}$-space as
\begin{equation}
\label{H_alternating}
H=\sum_{{\textbf k},\sigma}
\left(
\begin{matrix}
a^\dagger_{{\textbf k},\sigma} & b^\dagger_{{\textbf k},\sigma}
\end{matrix}
\right)
\left(
\begin{matrix}
V & \varphi^*({\textbf k}) \\
\varphi({\textbf k}) & -V
\end{matrix}
\right)
\left(
\begin{matrix}
a_{{\textbf k},\sigma} \\
b_{{\textbf k},\sigma}
\end{matrix}
\right)
\end{equation}
with
\begin{equation}
\label{phi}
\varphi({\textbf k})=-t\biggr[ e^{ik_x a/\sqrt{3}}+ 2e^{-ik_x a/(2\sqrt{3})}\cos\left(\frac{k_ya}{2}\right)\biggr].
\end{equation}
In Eq. (\ref{H_alternating}) $a^\dagger_{{\textbf k},\sigma}$($a_{{\textbf k},\sigma}$) creates(annihilates) an electron at sublattice $A$, while $b^\dagger_{{\textbf k},\sigma}(b_{{\textbf k},\sigma}$) creates(annihilates) an electron at sublattice $B$, with momentum ${\textbf k}$ and spin $\sigma$. $\varphi({\textbf k})$ is the nearest-neighbor hopping energy dispersion, with $t$ being the nearest neighbor hopping integral and $a$ being the lattice constant.

Since the Hamiltonian contains no spin-dependent interactions, we can define the corresponding Green's function matrix as
\begin{equation}
\label{GR_toy}
[G({\textbf k},z)]=
\left(
\begin{matrix}
z - V & - \varphi^*({\textbf k})\\
- \varphi({\textbf k}) & z + V
\end{matrix}
\right)^{-1},
\end{equation}
with $z=\omega \pm i0^+$, and $\omega$ real frequency variable. Here, we set $\hbar=1$ such that frequency and energy have the same unit. We can then use this Green's function matrix to compute the DOS
\begin{equation}
\label{DOS_toy}
\rho(\omega)= \sum_{{\textbf k},\sigma} -\frac{1}{\pi} {\rm Im} {\rm Tr} [G({\textbf k},\omega+i0^+)],
\end{equation}
and the optical conductivity tensor derived from the Kubo formula,
as previously used in Refs. [\onlinecite{Majidi_PRB_2011,Majidi_PRB_2013}],
\begin{eqnarray}
\label{opt_cond}
\sigma_{\alpha \beta}(\omega) \propto
\int d\nu \biggr( \frac{f(\nu,\mu,T)-f(\nu+\omega,\mu,T)}{\omega} \biggr) ~~~~~~~~~~
\nonumber  \\
~~~~  \times \sum_{{\textbf k}}
~ {\rm Tr} [v_{\alpha}({\textbf k})][A({\textbf k},\nu)]
[v_{\beta}({\textbf k})][A({\textbf k},\nu+\omega)],
\end{eqnarray}
with
$[v_{\lambda}({\textbf k})]=
\partial[H({\textbf k})]/\partial k_{\lambda}$
as the Cartesian component of the velocity matrix,
\begin{equation}
[A({\textbf k},\nu)] =\frac{i}{2\pi} \big\{[G({\textbf k},\nu+i0^+)]-
[G({\textbf k},\nu-i0^+)]\big\},
\end{equation}
as the spectral function matrix,
$f(\nu,\mu,T)$ as the Fermi distribution function, $\omega$ and $\nu$ as the real frequency variables, $\mu$ as the chemical potential, and $T$ as the temperature.
Here, the transverse component $\sigma_{xy}(\omega)=\sigma_{yx}(\omega)\approx 0$, hence we shall address only the longitudinal component $\sigma_1(\omega)\equiv\sigma_{xx}(\omega)\approx\sigma_{yy}(\omega)$. Furthermore, we are interested in the ground-state properties, i.e., at $T=0$.
\begin{figure}
\includegraphics[width=3.2in]{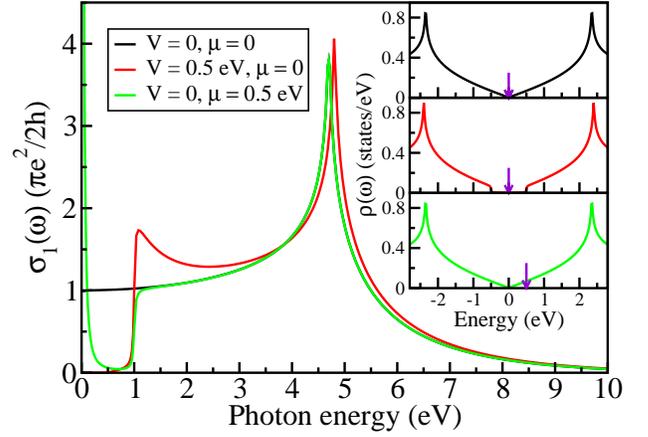}
\caption{(Color online) Comparison between optical conductivity of (a) bare graphene, (b) graphene with an alternating potential, and (c) charged
graphene. The magnitude of the alternating potential and the chemical potential are taken to be as follows: $V=\mu=0$ for the bare graphene, $V=0.5$ eV, $\mu=0$ for graphene with alternating potential, and $V=0, \mu=0.5$ eV for charged graphene, respectively. The insets show the DOS for the three cases where the small vertical arrowed line in each inset indicates the position of the chemical potential.}
\label{bare_gap_Pauli_blocking}
\end{figure}

Let us compare the results for bare graphene ($V=\mu=0$), graphene with alternating potential ($V=0.5$ eV, $\mu=0$), and charged graphene ($V=0$, $\mu=0.5$ eV), that are shown in Fig. \ref{bare_gap_Pauli_blocking}. $\sigma_1(\omega)$ for the bare graphene [see the black curve (a) in the main panel] has a finite value at the dc limit which is the universal value $\sigma_0=\pi e^2/2h$ and has a peak associated with the transition between the two Van Hove singlarity points of the DOS. Here, we take $t=2.35$ eV such that the position of the peak agrees with the experimental data of $\sigma_1(\omega)$ we will address later. As we introduce the alternating potential, the gap of size $\Delta\approx 2V=1$ eV forms in the DOS [see inset (b)]. Corresponding to the gap formation in the DOS, $\sigma_1(\omega)$ reveals a complete loss of spectral weight from $\omega=0$ to $\omega\approx\Delta$ [see red curve (b) in the main panel]. To compensate the $f$-sum rule, this loss of spectral weight is accompanied with the appearance of a "hump" at the gap edge. 

Now, let us see the situation for the charged graphene. Since the electrons are assummed noninteracting, shifting the chemical potential ($\mu$) does not alter the profile of DOS compared to that of the bare graphene. However, the shift of the position of $\mu$ significantly affects the profile of $\sigma_1(\omega)$. Here we see a similarity between the profile of $\sigma_1(\omega)$ for graphene with a gap and charged graphene without a gap [compare curves (b) and (c) of $\sigma_1(\omega)$] in the energy region $0<\omega\leq 1$ eV but not quite close to 0. In this region the green curve (c) also shows a vanishing spectral weight akin to the formation of gap. This is known to be due to the Pauli blocking effect \cite{Grigorenko_Nat_Phot_2012}. It is important, however, to note the difference. Charged graphene has a finite DOS at $\mu$, hence it is a metal rather than an insulator. This metallic characteristics is revealed by the formation of a sharp Drude peak at $\omega=0$ (dc conductivity). Therefore, an optical probe of $\sigma_1(\omega)$ not close enough to the dc limit may not be able to distinguish whether the system is gapped or only manifesting the Pauli blocking effect.

Note that if $\mu$ is too small, away from the Drude response $\sigma_1(\omega)$ may not completely vanish before it rises back. Such a situation reveals in graphene under a gate voltage as high as 71 volts \cite{Li_Nat_Phys_2008}, from which we may infer that $2\mu\approx5000~{\rm cm}^{-1}$, or $\mu\approx0.3$ eV. This situation may be explained as the following. In a Pauli-blocked system of graphene, the Drude peak has a width, and hence a tail, which samples the finite values of the density of states closely below and above $\mu$.
If $\mu$ is not big enough this tail effect would produce some residual conductivity in the energy region between 0 and $2\mu$.
Theoretical studies on graphene with a shifted chemical potential that incorporate interactions among electrons and between electrons and other quasi-particles suggest additional broadening of the Drude peak and renormalization of the gap edge associated with the interband transition \cite{Grushin_PRB_2009, Carbotte_PRB_2012}. These factors, in turn, enhance the residual conductivity mentioned above.
\\

\begin{center}
\textbf{III. MODEL FOR GRAPHENE WITH ADATOMS}
\end{center}

In what follows we discuss a model for graphene with magnetic adatoms. Despite the fact that adatoms on graphene may tend to form clusters \cite{Lehtinen_PRL_2003} or ordered states \cite{Cheianov_PRB_2009}, for simplicity we assume that they are distributed randomly over the graphene sheet and each adatom rests on top of a carbon atom. Suppose we have $x$ portion of adatoms out of $N$ lattice sites. Thus, statistically we expect to have $x(1-x)$ portion of supercells containing only an adatom at sublattice $A$, the $x(1-x)$ portion containing only an adatom at sublattice $B$, the $x^2$ portion containing adatoms at both sublattices $A$ and $B$, and the $(1-x)^2$ portion with no adatoms [see the illustration in Fig. \ref{model_structure}].

Considering that each adatom carries a magnetic moment corresponding to its spin, we introduce local magnetic interactions between the carrier spins and the spins of the adatoms such that the full Hamiltonian reads
\begin{eqnarray}
\label{H_fullmodel}
H=\sum_{{\textbf k}}
\Psi^\dagger_{\textbf k}
\left(
\begin{matrix}
0 & \varphi^*({\textbf k}) & 0 & 0   \\
\varphi({\textbf k}) & 0 & 0 & 0 \\
0 & 0 & 0 & \varphi^*({\textbf k})    \\
0 & 0 & \varphi({\textbf k}) & 0
\end{matrix}
\right)
\Psi({\textbf k}) \nonumber \\
- J_H \biggr( \sum_i {\textbf S}_{A i}\cdot {\textbf s}_{A i}
+ \sum_j {\textbf S}_{B j}\cdot {\textbf s}_{B j}\biggr),
\end{eqnarray}
where we have used spinor notations,
\begin{eqnarray}
\Psi^\dagger ({\textbf k})=
(
a^\dagger_{{\textbf k},\uparrow} ~  b^\dagger_{{\textbf k},\uparrow} ~  a^\dagger_{{\textbf k},\downarrow}
~ b^\dagger_{{\textbf k},\downarrow}
) ~~
{\rm and} ~~
\Psi({\textbf k})=
\left(
\begin{matrix}
a_{{\textbf k},\uparrow} \\
b_{{\textbf k},\uparrow} \\
a_{{\textbf k},\downarrow} \\
b_{{\textbf k},\downarrow}
\end{matrix}
\right). \nonumber 
\end{eqnarray}

\noindent In the second line of Eq. (\ref{H_fullmodel}) ${\textbf S}_{A i} ({\textbf S}_{B j})$ denotes the spin of an adatom on top of site $i(j)$ of sublattice $A(B)$, whereas ${\textbf s}_{A i} ({\textbf s}_{B j})$  denotes the spin of an electron correspondingly. $J_H$ is Hund's coupling between an electron spin and
spin of an adatom at the same site. It is customary to treat the local spins in the above Hamiltonian classically. Such a problem would then be
easily solved within the dynamical mean-field theory (DMFT) \cite{DMFTGeorgesRMP96,Furukawa94} as similarly performed in Ref. [\onlinecite{Majidi_PRB_2008}].

To start the DMFT algorithm we define the Green's function matrix through the Dyson equation,
\begin{equation}
\label{Dyson}
[G({\textbf k},z)]^{-1}=[G_0({\textbf k},z)]^{-1}-[\Sigma(z)],
\end{equation}
with
\begin{equation}
\label{G0}
[G_0({\textbf k},z)]=
\left(
\begin{matrix}
z  & - \varphi^*({\textbf k}) & 0 & 0\\
- \varphi({\textbf k}) & z  & 0 & 0 \\
0 & 0 & z  & - \varphi^*({\textbf k})\\
0 & 0 & - \varphi({\textbf k}) & z
\end{matrix}
\right)^{-1}
\end{equation}
the unperturbed Green's function matrix, and $[\Sigma(z)]$ as some initial guess of self-energy matrix.
Then, we average $[G({\textbf k},z)]$ over the first Brillouin zone of graphene,
$[G(z)]=\sum_{\textbf k} [G({\textbf k},z)]$.
The true $[\Sigma(z)]$ is to be solved through a self-consistency process.

The mean-field Green's function matrix is extracted through
$[{\cal G}(z)]=[ [G(z)]^{-1}+[\Sigma(z)]]^{-1}$. This is then used to compute
the local interacting Green's function matrix for a supercell with: an adatom at sublattice $A$ only, $[G_A(z,\theta_A,\phi_A)]$;
an adatom at sublattice $B$ only, $[G_B(z,\theta_B,\phi_B)]$;
and adatoms at both sublattices $A$ and $B$, $[G_{AB}(z,\theta_A,\phi_A,\theta_B,\phi_B)]$. Each inverse of these
local interacting Green's function matrices is obtained by substracting its corresponding local interaction matrix,
\begin{eqnarray}
\Sigma_{A}=- \frac{J_HS}{2}
\left(
\begin{matrix}
\cos\theta_A  & 0 & \sin\theta_Ae^{-i\phi_A} & 0\\
0 & 0  &  0 & 0\\
\sin\theta_Ae^{i\phi_A} & 0 & -\cos\theta_A  & 0\\
0 & 0 & 0 & 0\\
\end{matrix}
\right), \nonumber \\
\Sigma_{B}=- \frac{J_HS}{2}
\left(
\begin{matrix}
0  & 0 & 0 & 0\\
0 & \cos\theta_B  &  0 & \sin\theta_Be^{-i\phi_B}\\
0 & 0 & 0  & 0\\
0 & \sin\theta_Be^{i\phi_B} & 0 & -\cos\theta_B\\
\end{matrix}
\right),  \nonumber
\end{eqnarray}
\begin{widetext}
\noindent and
\begin{eqnarray}
\Sigma_{AB}=- \frac{J_HS}{2}
\left(
\begin{matrix}
\cos\theta_A  & 0 & \sin\theta_Ae^{-i\phi_A} & 0\\
0 & \cos\theta_B  &  0 & \sin\theta_Be^{-i\phi_B}\\
\sin\theta_Ae^{i\phi_A} & 0 & -\cos\theta_A  & 0\\
0 & \sin\theta_Be^{i\phi_B} & 0 & -\cos\theta_B\\
\end{matrix}
\right) \nonumber
\end{eqnarray}
from $[{\cal G}(z)]^{-1}$.
The three local interacting Green's function matrices, corresponding to the supercells with different adatom occupations are to be averaged over all possible directions of adatom spins ${\textbf S}_A$ and  ${\textbf S}_B$ in their corresponding solid angles $\Omega_A$ and $\Omega_B$,
\begin{eqnarray}
&&[G_A(z)]=\int d\Omega_A P_A(\theta_A,\phi_A)[G_A(z,\theta_A,\phi_A)],  \nonumber \\
&& [G_B(z)]=\int d\Omega_B P_B(\theta_B,\phi_B)[G_B(z,\theta_B,\phi_B)],  \\
&&[G_{AB}(z)]=\int d\Omega_A d\Omega_B P_{AB}(\theta_A,\phi_A,\theta_B,\phi_B) [G_{AB}(z,\theta_A,\phi_A,\theta_B,\phi_B)]. \nonumber
\end{eqnarray}

Here, $P_A(\theta_A,\phi_A)$ and $P_B(\theta_B,\phi_B)$ are assumed constant for all angles since ${\textbf S}_A$ or
${\textbf S}_B$  standing alone in a supercell is assumed to be free to point to any direction. Whereas, $P_{AB}(\theta_A,\phi_A,\theta_B,\phi_B)$ depends on the assumed ground-state magnetic configuration of the adatoms, namely
\begin{eqnarray}
P_{AB}(\theta_A,\phi_A,\theta_B,\phi_B) \propto 
\begin{cases}
\delta(\theta_A)\delta(\theta_B-\pi)  & {\rm for ~ the ~ AFM ~ configuration,} \nonumber \\
\delta(\theta_A)\delta(\theta_B) & {\rm for ~ the ~ FM ~ configuration,} \nonumber \\
{\rm constant~} & {\rm for ~ the ~PM ~ configuration.}
\end{cases}
\end{eqnarray} 
\end{widetext}

The average interacting Green's function matrix is then obtained by further averaging over supercells with different adatom occupations,
\begin{eqnarray}
[G(z)]_{ave}=x(1-x)[G_A(z)]+x(1-x)[G_B(z)] \nonumber \\
+ x^2[G_{AB}(z)]+(1-x)^2[{\cal G}(z)].
\end{eqnarray}
Finally, the corrected self-energy matrix is extracted from the Dyson equation $[\Sigma(z)]=[{\cal G}(z)]^{-1}-[G(z)]_{ave}^{-1}$, which is then used to start the new iteration by feeding it back into Eq. (\ref{Dyson}). This self-consistency process is continued until $[\Sigma(z)]$
is converged. Once the convergency is achieved, we compute the DOS and $\sigma_1(\omega)$ according to the definions given by Eq. ({\ref{DOS_toy}) and Eq. ({\ref{opt_cond}), respectively, except that now the summation over the spin index is not needed as the spin dependence has been taken explicitly into the definition of the matrices.
\\

\begin{center}
\textbf{IV. DISCUSSION ON VARIOUS MAGNETIC GROUND STATES}
\end{center}

\begin{figure}
\includegraphics[width=3.2in]{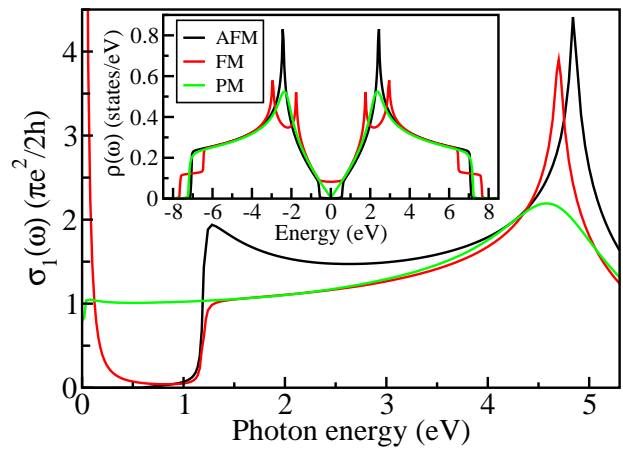}
\caption{(Color online) Comparison of optical conductivities of graphenes with 100\% adatom coverage for various possible magnetic configurations.
Main panel shows the optical conductiviy for (a) antiferromagnetic (AFM), (b) ferromagnetic (FM), and (c) paramagnetic (PM) ground-states, respectively.
The insets show their corresponding DOS profiles.}
\label{AFM_FM_PM}
\end{figure}

Figure \ref{AFM_FM_PM} displays our calculation results of optical conductivities of graphenes with 100\% adatom coverage ($x=1$) for various possible magnetic configurations.
Note that, despite the fact that the calculations we perform are for the ground states, i.e., $T=0$, the results should still be relevant to compare with finite temperature experimental data as long as the temperature is well below any magnetic ordering transition temperature.
For these calculations we choose the parameter values $S_A=S_B=1/2$, $J_H=1.2$ eV, and $t=2.35$ eV. The main panel shows the distinction among $\sigma_1(\omega)$'s for AFM, FM, and PM configurations. From the DOS profiles shown in the insets, it is clearly seen that a gap forms only for the AFM ground state. We also compute the total electronic energy, $\int_{-\infty}^\mu \omega \rho(\omega)d\omega$ for each of the three possible ground-state configurations and confirm that the AFM configuration gives the lowest energy. The physical concept of the formation of the gap in the AFM configuration is essentially that the AFM configuration induces alternating potentials as discussed previously above.

At this point, we would like to address the experimental data of fluorinated graphene presented in Ref. [\onlinecite{Nair_Nat_Phys_2012}] in connection with our results. Because the vanishing contribution of the most magnetic moments is due to clustering of the fluorine adatoms, we argue that every pair of neighboring magnetic moments inside the cluster align antiferromagnetically, whereas only the magnetic moments at the edge have the freedom to change their orientations. To verify our prediction, an optical conductivity measurement is necessary for which we predict a profile of $\sigma_1(\omega)$, such as the black curve in Fig. \ref{AFM_FM_PM} would result.
\\

\begin{center}
\textbf{V. ANALYSIS OF OPTICAL CONDUCTIVITY WITH SIMILAR GAPLIKE STRUCTURES}
\end{center} 

\begin{figure}
\includegraphics[width=3.2in]{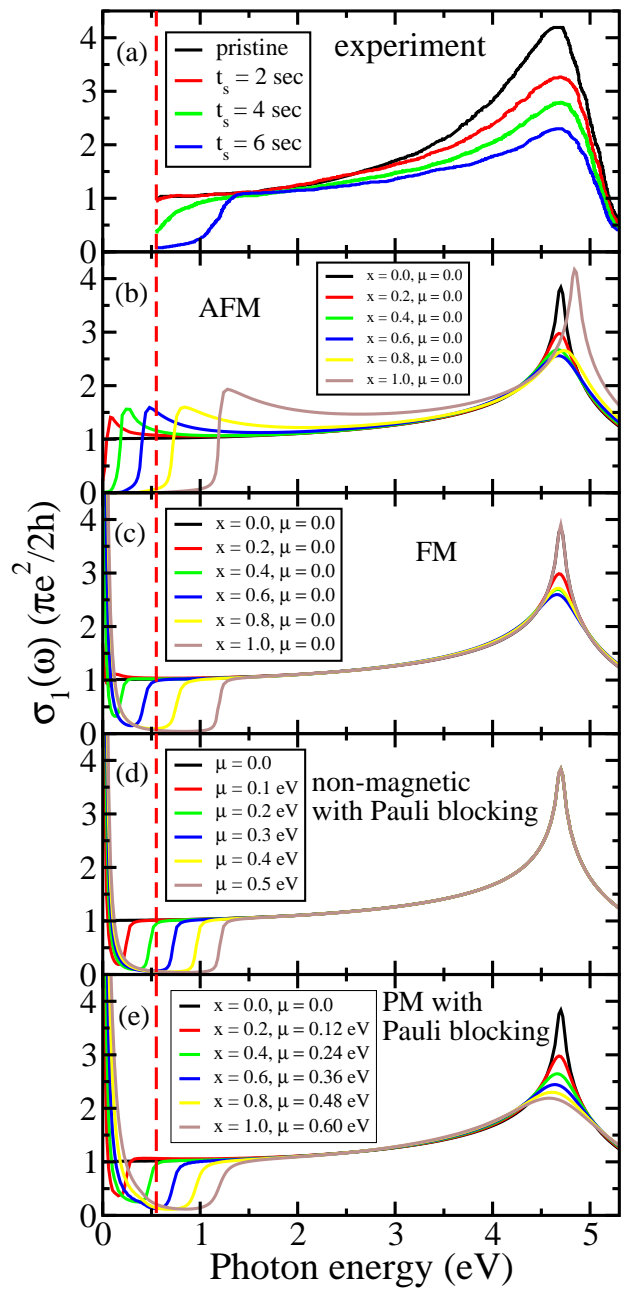}
\caption{(Color online) Comparison of  $\sigma_1(\omega)$ between experimental data and calculation results. (a) The experimental data of oxygenated graphene for various times of oxygen exposure (replotted from Ref. \onlinecite{Santoso_PRB_2014}). (b) and (c) Graphene systems with FM- and AFM- ordered adatoms, respectively, for various adatom concentrations ($x$). (d) Graphene with nonmagnetic adatoms for various chemical potential values ($\mu$). (e) Graphene with magnetic adatoms of the PM configuration with $\mu$ increasing proportionally to $x$.
The vertical red dashed line is to mark the energy value below which the experimental data are not available for comparison.}
\label{comp_theory_exp}
\end{figure}

In the following we analyze $\sigma_1(\omega)$ for four different circumstances leading to similar gaplike structures. Our aim is to propose a means to predict the magneticity and the possible orderings of the adatoms on graphene solely from the experimental optical conductivity data. As an example, we apply this analysis to the recently reported experimental data of oxygenated graphene \cite{Santoso_PRB_2014} [replotted in panel (a) of Fig. \ref{comp_theory_exp}]. Since, unfortunately, the experimental data of $\sigma_1(\omega)$ for photon energy below 0.55 eV are not yet available, we focus on analyzing the behavior of $\sigma_1(\omega)$ within the available energy range ($\sim 0.55-5.3$ eV).
The result of this analysis may later be verified when the low energy data have become available.
Note that our assumption on the possibility that oxygen adatoms possess magnetic moments is based on several experimental hints \cite{Elfimov_PRL_2007,Herng_PRL_2010,Rusydi_PTRSA_2012} suggesting that oxygen can be magnetic in certain environments.

The experimental data of $\sigma_1(\omega)$ displayed in Fig. \ref{comp_theory_exp} (a) show that a gaplike structure gradually forms, whereas the 4.7-eV peak monotonically decreases its intensity as the system adsorbs more oxygen adatoms.
To compare with our calculations, first, we assume that the adatoms are magnetic and form an AFM order and explore the system for various adatom concentrations ($x$) as shown in panel (b). As $x$ is increased, a real gap emerges accompanied by the appearance of a hump at the gap edge. The 4.7-eV peak intensity decreases as $x$ is increased up to 0.6 and increases back accompanied with a blue shift for a further increase in $x$. Apparently, the appearance of a hump and the nonmonotonic change in intensity of the 4.7-eV peak do not agree with those observed in the experimental data.

Second, as shown in panel (c), we explore the system with FM oder for various $x$'s as performed for the AFM case. As $x$ is increased, a gaplike structure with a Drude peak forms. Apart from the Drude peak, this behavior looks similar to that in the experimental data. The 4.7-eV peak also changes its intensity as $x$ is increased up to 0.6, but it increases back as $x$ is increased further. Again, this nonmonotonic change in intensity of the 4.7-eV peak does not agree with that observed in the experimental data.

Third, we consider nonmagnetic adatoms and assume that they supply some extra charge to graphene, such that the chemical potential ($\mu$) is shifted  proportionally to $x$. As shown in panel (d), shifting $\mu$ away from 0 leads to the formation of a gaplike structure with a Drude peak in the low-energy region. As far as for the energy region above 0.55 eV, this behavior looks similar to that in the experimental data. However, we notice that the 4.7-eV peak does not change as $\mu$ changes, contradictory to the experimental data.

Last, we consider graphene with magnetic adatoms in PM configuration [see panel (e)]. Here we also assume that $\mu$ varies proportionally to $x$. By increasing $x$, hence also $\mu$, we generate a gaplike structure with a Drude peak. Apart from the Drude peak, remarkably we see that the 4.7-eV peak monotonically decreases as $x$ increases, consistent with that observed in the experimental data. Thus, within the energy range of 0.55-5.3 eV,  $\sigma_1(\omega)$ of the oxygenated graphene seems to resemble that of graphene with the adatoms of the PM configuration.

Now let us elaborate further our results in general. We have seen that a hump structure appears in $\sigma_1(\omega)$ whenever the system forms a gap, suggesting that the hump may be a signature of the presence of a gap in a graphene sample. On the other hand, the absence of a hump of a gaplike structure in  $\sigma_1(\omega)$ may be the signature of the presence of a Drude peak or the Pauli-blocking effect. Meanwhile, by inspecting whether or not the 4.7-eV peak changes its intensity as a function of adatom concentration, we may predict that the adatoms posses magnetic moments if the peak intensity changes with variation of adatom concentration, otherwise, the adatoms are nonmagnetic. Interestingly, how the 4.7-eV peak changes its intensity as a function of adatom concentration may signify a possible magnetic ordering of the adatoms. Namely, the adatoms may form: AFM order if the 4.7-eV peak shows a nonmonotonic change in its intensity, accompanied by a blue shift; FM order if the 4.7-eV peak shows a nonmonotonic change in its intensity, without a blue shift; or PM configuration if the 4.7-eV peak shows a monotonic dicrease in its intensity; as the adatom concentration is increased. Such connections between the optical conductivity data and the  magneticity and the ordering of the adatoms, if true, may provide a promising means to deduce the magnetic characteristics of adatoms on graphene in case magnetic probes are not available. We therefore suggest that it be crucial to verify these connections by experiments with magnetic probes and/or optical conductivity down to very low energy region close to the dc limit.

Lastly, we comment on the little dissimilarity of our results as to compare with the experimental data of oxygenated graphene. Namely, the 4.7 eV peak appears to be much broader in the experimental data than that obtained in our calculations. We conjecture that this is due to some factors not taken into account in our model, possibly the effects of a substrate, or interplay between electron-electron Coulomb interactions and interactions between electrons spins and the spins of the adatoms. These demand further theoretical studies.\\

\begin{center}
\textbf{V. CONCLUSION}
\end{center}

We have presented our theoretical study of graphene with magnetic and nonmagnetic adatoms. We start with an introductory study of systems of graphene with alternating potentials and a bare graphene with a shifted chemical potential. We demonstrate the optical conductivities corresponding to these systems. Both systems posses similar gaplike structures in $\sigma_1(\omega)$, but they differ in that only graphene with the alternating potential reveals a real gap, whereas graphene with the shifted chemical potential reveals a sharp Drude peak. Furthermore, we study a model for graphene with adatoms with magnetic interactions between electron spins and spins of the adatoms. 
We demonstrate the different profiles of $\sigma_1(\omega)$ corresponding to AFM, FM, and PM adatom configurations. We show that only graphene with AFM adatom configurations possesses a gap, and this configuration gives the lowest electronic energy.
Furthermore, we analyze four different circumstances leading to similar gaplike structures in $\sigma_1(\omega)$ and propose connections between characteristics revealed in $\sigma_1(\omega)$ and the magneticity and the spin ordering of the adatoms. We compare our results with
the experimental data of oxygenated graphene from Ref. \onlinecite{Santoso_PRB_2014}, from which we conjecture that the adatoms are magnetic with the PM configuration. Our results show a path to modify optical properties as well as band gap of graphene via magnetic adatoms.
\\

\begin{center}
\textbf{ACKNOWLEDGEMENT}
\end{center}

This work is supported by Singapore National Research Foundation under its Competitive Research Funding (NRF-CRP 8-2011-06 and NRF2008NRF-CRP002024), MOE-AcRF Tier-2 (MOE2010-T2-2-121), NUS-YIA, and FRC, and in part, by Hibah PUPT from Indonesian Directorate General for Higher Education. We acknowledge the CSE-NUS computing centre for providing facilities for our numerical calculations.


\begin{thebibliography}{99}
\bibitem{Schedin_Nat_Mat_2007} F. Schedin, A. K. Geim, S. V. Morozov, E. W. Hill, P. Blake, M. I. Katsnelson, and K. S. Novoselov, Nature Mater. {\textbf 6}, 652 (2007).
\bibitem{Chen_Nat_Phys_2008} J.-H. Chen, C. Jang, S. Adam, M. S. Fuhrer, E. D. Williams, and M. Ishigami, Nat. Phys. {\textbf 4}, 377 (2008).
\bibitem{Zhou_PRL_2008} S. Y. Zhou, D. A. Siegel, A. V. Fedorov, and A. Lanzara, Phys. Rev. Lett. {\textbf 101}, 086402 (2008).
\bibitem{Elias_Sci_2009} D. C. Elias, R. R. Nair, T. M. G. Mohiuddin, S. V. Morozov, P. Blake, M. P. Halsall, A. C. Ferrari, D. W. Boukhvalov, M. I. Katsnelson, A. K. Geim, and K. S. Novoselov, Science {\textbf 323}, 610 (2009).
\bibitem{CastroNeto_SSC_2009} A.H. Castro Neto, V.N. Kotov, J. Nilsson, V.M. Pereira, N.M.R. Peres, B. Uchoa, Solid State Commun. {\textbf 149} 1094 (2009).
\bibitem{Nair_Nat_Phys_2012} R. R. Nair, M. Sepioni, I-Ling Tsai, O. Lehtinen, J. Keinonen, A. V. Krasheninnikov, T. Thomson,
A. K. Geim, and I. V. Grigorieva, Nat. Phys. {\textbf 8}, 199 (2012).
\bibitem{Kan_Nano_2008} E. Kan, Z. Li, and J. Yang, Nano {\textbf 3}, 443 (2008).
\bibitem{Brey_PRL_2007} L. Brey, H. A. Fertig, and S. Das Sarma, Phys. Rev. Lett. {\textbf 99}, 116802 (2007).
\bibitem{Saremi_PRB_2007} S. Saremi, Phys. Rev. B {\textbf 76}, 184430 (2007).
\bibitem{Santoso_PRB_2014} I. Santoso, R. S. Singh, P. K. Gogoi, T. C. Asmara, D. Wei, W. Chen, A. T. S. Wee, V. M. Pereira, and A. Rusydi, Phys. Rev. B {\textbf 89}, 075134 (2014).
\bibitem{Kotov_RMP_2012} V. N. Kotov, B. Uchoa, V. M. Pereira, F. Guinea, and A. H. Castro Neto, Rev. Mod. Phys. {\textbf 84}, 1067 (2012).
\bibitem{Majidi_PRB_2011} M. A. Majidi, H. Su, Y. P. Feng, M. R\"{u}bhausen, and A. Rusydi, Phys. Rev. B {\textbf 84}, 075136 (2011).
\bibitem{Majidi_PRB_2013} M. A. Majidi, E. Thoeng, P. K. Gogoi, F. Wendt, S. H. Wang, I. Santoso, T. C. Asmara, I. P. Handayani, P. H. M. van Loosdrecht, A. A. Nugroho, M. R\"{u}bhausen, and A. Rusydi, Phys. Rev. B {\textbf 87}, 235135 (2013).
\bibitem{Grigorenko_Nat_Phot_2012} A. N. Grigorenko, M. Polini, and K. S. Novoselov, Nature Photon. {\textbf 6}, 749 (2012).
\bibitem{Li_Nat_Phys_2008} Z. Q. Li, E. A. Henriksen, Z. Jiang, Z. Hao, M. C. Martin, P. Kim, H. L. Stormer, and D. N. Basov, Nat. Phys. {\textbf 4}, 533, (2008).
\bibitem{Grushin_PRB_2009} A. G. Grushin, B. Valenzuela, and M. A. H. Vozmediano, Phys. Rev. B {\textbf 80}, 155417 (2009).
\bibitem{Carbotte_PRB_2012} J. P. Carbotte, J. P. F. LeBlanc, and E. J. Nicol, Phys. Rev. B {\textbf 85}, 201411(R) (2012).
\bibitem{Lehtinen_PRL_2003} P. O. Lehtinen, A. S. Foster, A. Ayuela, A. Krasheninnikov, K. Nordlund, and R. M. Nieminen, Phys. Rev. Lett. {\textbf 91}, 017202 (2003).
\bibitem{Cheianov_PRB_2009} V. V. Cheianov, O. Sylju{\aa}sen, B. L. Altshuler, V. Fal'ko, Phys. Rev. B {\textbf 80}, 233409 (2009).
\bibitem{DMFTGeorgesRMP96} A. Georges, G. Kotliar, W. Krauth and M. Rozenberg, Rev. Mod. Phys. {\textbf 68}, 13 (1996).
\bibitem{Furukawa94} N. Furukawa, J. Phys. Soc. Japan {\textbf 63}, 3214 (1994) and
Proc. Conference on Physics of Manganites (1998) (available at http://xxx.lanl.gov/abs/cond-mat/9812066).
\bibitem{Majidi_PRB_2008} M. A. Majidi, M. Jarrell, J. Moreno, R. Fishman, and K. Aryanpour, Phys. Rev. B {\textbf 74}, 115205 (2006).
\bibitem{Elfimov_PRL_2007} I. S. Elfimov, A. Rusydi, S. I. Csiszar, Z. Hu, H. H. Hsieh, H.-J. Lin, C. T. Chen, R. Liang, and G. A. Sawatzky, Phys. Rev. Lett. {\textbf 98}, 137202 (2007).
\bibitem{Herng_PRL_2010} T.S. Herng, D.-C. Qi, T. Berlijn, J.B. Yi, K.S. Yang, Y. Dai, Y.P. Feng, I. Santoso, C. Sánchez-Hanke, X.Y. Gao, Andrew T.S. Wee, W. Ku, J. Ding, and A. Rusydi, Phys. Rev. Lett. {\textbf 105}, 207201 (2010).
\bibitem{Rusydi_PTRSA_2012} A. Rusydi, S. Dhar, A. Roy Barman, Ariando, D.-C. Qi, M. Motapothula, J. B. Yi, I.Santoso, Y. P. Feng, K. Yang, Y. Dai, N. L. Yakovlev, J. Ding, A. T. S. Wee, G.Neuber, M. B. H. Breese, M. Ruebhausen, H. Hilgenkamp and T. Venkatesan, Phil. Trans. R. Soc. A {\textbf 370}, 4927-4943 (2012).

\end{thebibliography}
\end{document}